\documentclass[twocolumn,twoside,preprintnumbers,amsmath,amssymb]{revtex4}
\usepackage{epsfig}
\usepackage{graphicx}
\newcommand{\bra}[1]{\left\langle #1 \right|}
\newcommand{\ket}[1]{\left| #1 \right\rangle}
\newcommand{\braket}[2]{\left\langle #1 | #2 \right\rangle}
\begin{document}
\title{Exchange Energy in Coupled Quantum Dots}
\author{H. E. Caicedo-Ortiz \thanks{hecaicedo@unicauca.edu.co} \space and \space S. T. Perez-Merchancano\thanks{sperez@unicauca.edu.co}\\
{\small\it Departamento de F\'{i}sica, Universidad del Cauca,
Calle 5 \# 4-70, Campus Tulc\'{a}n, Popay\'{a}n, Colombia }}
\begin{abstract}
In this work, the exchange energy $J$ for a system of two
laterally-coupled quantum dots, each one with an electron, is
calculated analytically and in a detailed form, considering them
as hydrogen-like atoms, under the Heitler-London approach. The
atomic orbitals, associated to each quantum dot, are obtained from
translation relations, as functions of the Fock-Darwin states. Our
results agree with the reported ones by Burkard, Loss and
DiVincenzo in their model of quantum gates based on quantum dots,
as well as with some recent experimental reports.
\end{abstract}
\maketitle
\baselineskip=12pt
\section{Introduction}

In the last decade, a great interest in quantum dots\cite{jacak}
has arouse, due to their potential use as hardware for the
implementation of a scalable quantum computer\cite{loss, sherwin}.
In this scheme, the electron spin in these quantum dots is used as
the basic element for the transport of information (qubit).
Considering the fact that 1-$qubit$ and 2-$qubit$ gates are
sufficient to make any quantum algorithm \cite{barenco,
divincenzo}, a quantum computing device, based on quantum dots,
must have a mechanism by which two specific qubits could be
entangled to produce a fundamental 2-qubit quantum gate, such as
the Controlled-NOT gate $XOR$\cite{loss}. This process is achieved
through single qubits rotations and an adequate switching of
exchange energy $J(t)$ between the electronic spins $\mathbf{S_1}$
and $\mathbf{S_2}$ described by the Heisemberg spin exchange
hamiltonian for a system of two laterally quantum dots, under the
influence of a magnetic field, perpendicular to their surface.
This is the reason for the studies of quantum gates with coupled
quantum dots are reduced to get experimental control of the single
qubits rotations\cite{cerletti} and  the exchange energy. In this
work, considering the Heitler-London\cite{heitler} approach, we
present the process to obtain an expression of $J$ for this system
as a function of parameters that allow its experimental control,
with a detailed description which is not available in the work of
Burkard\cite{burkard}.

\section{Theory model of two laterally coupled quantum dots}

Let us consider a system of two laterally coupled quantum dots,
each one with an electron, constituted by electrical gating of
two-dimensional electron gas (2DEG), under the action of a
$z$-axis parallel magnetic field $\mathbf{B}$, and an electric
field $\mathbf{E}$ in $x$-direction. Its physical representation
is given by
\begin{equation}
\label{eq1}
H_t = H_1  + H_2  + V_{12}  + W,
\end{equation}

where $H_j$ is the Hamiltonian for the $1$ and $2$ quantum dot,
and this is

\begin{equation}
\label{eq2}
H_j  = \frac{1}{{2\mu }}\left[ {\mathbf{p}_j  + \frac{q_e}{c} \mathbf{A (r)}} \right]^2  + eEx_j  + \frac{\omega _0 ^2\mu}{2} \left[ {(x_j  \pm a)^2  + y_j ^2 } \right],
\end{equation}

The Coulomb interaction is given by $V_{12}$ , and $W=W_1  + W_2$
represent a quartic potential for each quantum dot and it is
written as
\begin{equation}
\label{eq3}
W_i  = \frac{{\mu \omega _0 ^2 }}{2}\left[ {\frac{1}{{4a^2 }}(x_i ^2  - a)^2  - (x_i \pm a)^2 } \right]^2 ; \   \   \   \    i=1,2,
\end{equation}
This potential models the effect of tunneling between the two
quantum dots and its choice is motivated by experimental
evidence\cite{tarucha, kouwenhoven}. Considering a low-temperature
description, where the system is in a condition $kT < \hbar \omega
_0$, we can only assume the two lower orbital states of the
Hamiltonian $H_t$, which are singlet and triplet states. With
these conditions, and without considering Zeeman effect and
spin-orbit coupling, it is possible to translate this physical
picture into the Heisenberg spin Hamiltonian, which is
\begin{equation}
\label{eq4}
H_t = J\ \mathbf{S_1} \cdot \mathbf{S_2}
\end{equation}
This Hamiltonian is the scalar product between the spin operators
and the factor $J$, which is the exchange energy between the spin
triplet and singlet states\cite{mattis}.

\section{Wave Functions of Coupled Quantum Dots}
In order to apply the Heitler-London approach\cite{heitler}, it is
primarily necessary to determine the wave function of each quantum
dot, that constitutes our laterally coupled system, on which an
$xy$-axis electric field and a $z$-axis magnetic field act.

Considering symmetric translation operations on a quantum system,
through an scheme similar to the one used in the solution of a
charged harmonic oscillator in a uniform electric
field\cite{cohen}, it is possible to get the eigenfunctions of
(\ref{eq1})  for $j=1,2$. To do this, we write (\ref{eq1}) as a
momentum translation $p_{jy}$ of the Fock-Darwin hamiltonian plus
a constant depending of the electric field. The ground state of
the Hamiltonians $H_1$ and $H_2$ are,

\begin{equation}
\label{eq5} \phi_{0}^{(j)} = exp\left[-\Gamma{(j)}\right]
\sqrt{\frac{m\Omega}{\pi\hbar}}\
exp\left[-\frac{m\Omega}{2\hbar}\left(x_{\mp}^2+y^2\right)\right],
\end{equation}
for  j=1,2, where
\begin{equation}
\label{eq6}
\Gamma_{(j)}=\frac{i}{\hbar}\left(\frac{q_{e}^2BE}{2m\omega_{0}^2c}
\pm \frac{eBa}{2c}\right) y,
\end{equation}
with $x_{\pm}=x\pm a+\frac{q_{e}E}{m\omega{0}}$,  $q_e$ is the
electron charge,  $c$ is the light speed, $\mathbf{E}=(E,0,0)$,
$\mathbf{B}=(0,0,B)$,  $\Omega$ represents the  Fock-Darwin
frequency, $\omega_{0}$ the confinement frequency, and
$\mathbf{A(r)}=\frac{\mathbf{B}}{2}(-y,x,0)$ the vector potential.
These wave functions correspond to Fock-Darwin states\cite{jacak}
translated a certain amount of momentum $p_{jy}=p_{jy} +
\frac{q_{e}^2BE}{2m\omega_{0}^2c} \pm \frac{eBa}{2c}$.

\section{The Heitler-London approach}
A technique that allows to determine the exchange energy factor is
to adapt the Heitler-London method\cite{heitler} (also known as
valence orbit approximation) to our system, considering that it
behaves as a pair of hydrogen-like artificial atoms. The symmetric
(singlet) and antisymmetric (triplet) states are represented by

\begin{equation}
\label{eq7}
\ket{\psi_{\pm}}=\frac{\ket{A(1)B(2)}\pm\ket{A(2)B(1)}}{\sqrt{2(1\pm
S^2)}}.
\end{equation}

Applying the Heitler-London method on our system, we have
\begin{equation}
\label{eq8}\left. \begin{array}{l}
 A(j)= \phi_{0}^{(1)}  (x_j ,y_j)\\
 B(j) = \phi_{0}^{(2)} (x_j ,y_j )\\
 \end{array} \right\}{\rm{   }}j = 1,2.
\end{equation}

The parameter $S$ represents the overlapping between left and
right orbital in each dot.  This term is given by

\begin{equation}
\label{eq9}
S= \braket{\phi _0 ^{(2)}} {\phi _0 ^{(1)}},
\end{equation}

with
\begin{equation}
\label{eq10}
b = \frac{\Omega }{{\omega _0 }},\   \  \  \   d = a\sqrt{m\omega _0 /\hbar},
\end{equation}

where $a$ is half the distance between the centers of the dots,
$a_B=\sqrt{\hbar / m\omega _0 }$ is the effective Bohr radius of a
single isolated harmonics well, $d$ is the dimensionless distance,
and $b$ is a magnetic compression  factor of the quantum dots
orbitals.

\section{Exchange Energy}
According to the magnetism theory\cite{mattis}, the exchange
energy is represented by
\begin{equation}
\label{eq11} J= \epsilon_t - \epsilon_s = \left\langle {\psi _ - }
\right|H_t \left| {\psi _ -  } \right\rangle  - \left\langle {\psi
_ + } \right|H_t \left| {\psi _ +  } \right\rangle,
\end{equation}
where $ \epsilon_t $ represents the triplet energy,  $\epsilon_s$
singlet energy, and  $H_t $ is described by (\ref{eq1}).
Introducing this term in (\ref{eq11}), and regrouping common terms
we obtain
\begin{equation}
\label{eq12}
J = \frac{S^2 }{1 + S^4 }\bigl[ \Upsilon _1 - \frac{\Upsilon _2 } {S^2 } + \Upsilon _3  - \frac{\Upsilon _4 } {S^2 } +\Upsilon _5 \bigr],
\end{equation}
where
\begin{multline}
\label{eq13}
 \Upsilon _1  =
 \bra{A(1)}H_1\ket{A(1)}\braket{B(2)}{B(2)}\\
+\bra{B(2)}H_2\ket{B(2)}\braket{A(1)}{A(1)}\\
+\bra{B(1)}H_1\ket{B(1)}\braket{A(2)}{A(2)}\\
 \bra{A(2)}H_2\ket{A(2)}\braket{B(1)}{B(1)},
\end{multline}
\begin{multline}
\label{eq14}
 \Upsilon _2  =
\bra{A(1)}H_1\ket{B(1)}\braket{B(2)}{A(2)}\\
+\bra{B(2)}H_2\ket{A(2)}\braket{A(1)}{B(1)}\\
+\bra{B(1)}H_1\ket{A(1)}\braket{A(2)}{B(2)}\\
 \bra{A(2)}H_2\ket{B(2)}\braket{B(1)}{A(1)},
\end{multline}
\begin{multline}
\label{eq15}
\Upsilon _3  =
\bra{A(1) B{(2)}}V_{12}\ket{A(1) B{(2)}} \\
+\bra{A(2) B{(1)}}V_{12}\ket{A(2) B{(1)}},
\end{multline}
\begin{multline}
\label{eq16}
\Upsilon _4  = \bra{A(1) B{(2)}}V_{12}\ket{A(2) B{(1)}} \\
+\bra{A(2) B{(1)}}V_{12}\ket{A(1) B{(2)}},
\end{multline}
\begin{multline}
\label{eq17}
 \Upsilon _5  = \bra{A(1) B{(2)}}W_{1}+W_{2}\ket{A(1) B{(2)}} \\
+\bra{A(2)B{(1)}}W_{1}+W_{2}\ket{A(2) B{(1)}}\\
- \frac{1}{S^2}\bigl[\bra{A(2) B{(1)}}W_{1}+W_{2}\ket{A(1) B{(2)}} \\
+\bra{A(1) B{(2)}}W_{1}+W_{2}\ket{A(2) B{(1)}}\bigr].
\end{multline}
The solution of the term $\Upsilon_1$  (\ref{eq13}) does not have
higher difficult if we take that
$\braket{A(j)}{A(j)}=\braket{B(j)}{B(j)}=1$. For the next term
$\Upsilon_2$ (\ref{eq14}) it is easily to demonstrate that
$\braket{A(j)}{B(j)}=\braket{B(j)}{A(j)}=S$. The solutions of
$\Upsilon_3$ (\ref{eq15}) and $\Upsilon_4$ (\ref{eq16}) are
obtained using the center of mass and relative coordinates. The
next step is to make a change from cartesian coordinates to polar
coordinates. In this process, four kind of quadratures of many
special functions appear, which are resolved making use of the
expansions given in \cite{abramowitz} and \cite{gradshteyn}. In
order to determine $\Upsilon_5$  (\ref{eq17}) we write  $\bra{A(2)
B{(1)}}W_{1}+W_{2}\ket{A(2) B{(1)}}$ in terms of  $\bra{A(1)
B{(2)}}W_{1}+W_{2}\ket{A(1) B{(2)}}$ and the rest of terms are
calculated considering translation operations in $p_y$ and $x$.
Finally, replacing (\ref{eq9}), and the solutions of (\ref{eq13})
- (\ref{eq17}) in  (\ref{eq12}) we find
\begin{multline}
\label{eq18}
J(B,E,d)=\frac{\hbar\omega_0}{\sinh\bigl(2d^2\bigl[2b-\frac{1}{b}\bigr]\bigr)}
\Biggl[c\sqrt{b}\Biggr(e^{-bd^2}\mathbf{I}_{0}\left(bd^2\right)\\-e^{d^2\left(b-1/b\right)}\mathbf{I}_{0}\left(d^2\left[b-1/b\right]\right)\Biggr)+\frac{3}{4b}\left(1+bd^2\right)\\+\frac{3}{2}\frac{1}{d^2}\left(\frac{eEa}{\hbar
\omega_0}\right)^2\Biggr].
\end{multline}

This is the exchange energy, and as we can notice, it presents
dependence of external parameters such as $\mathbf{B}$,
$\mathbf{E}$ and $d$.

\section {Results}
Eq. (\ref{eq18}) describes the exchange energy $J$ and is
constituted by four terms. The first and second terms are result
of $\Upsilon_3$ and $\Upsilon_4$, in which the Coulomb interaction
$V_{12}$ acts. The third term, in spite of having a polynomial
behavior, avoids an abrupt decline that the two first terms offer.
Since there is a difference of sign between the first and second
term of Eq. (\ref{eq18}), there exists a value of
$\mathbf{B}=\mathbf{B^*}$ for which $J$ switches from positive to
negative.
\\
\begin{figure}[h]
\begin{center}
\includegraphics[angle=0,width=8.5 cm]{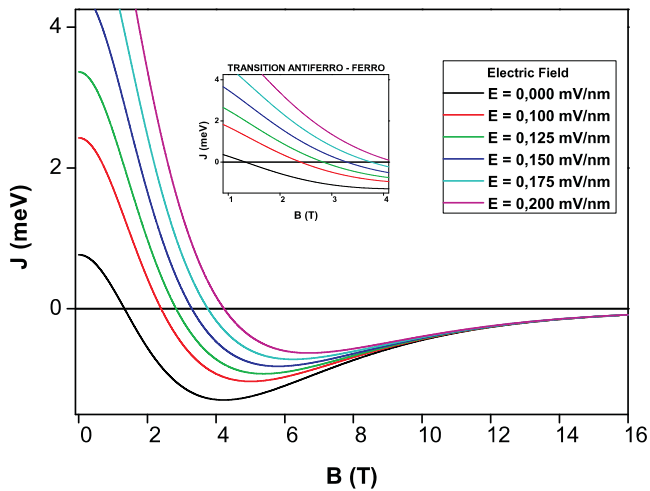}
\end{center}
\caption{Exchange energy $J(B)$ in $meV$ plotted against the
magnetic field for different electric fields, for $a=0.7 a_b$ and
$c=2.36.$ The most remarkable feature of $J$ is the change of its
sign from positive to negative.} \label{fig1}
\end{figure}

In Fig.\ref{fig1} we present the transition from antiferromagnetic
$(J>0)$ to ferromagnetic $(J<0)$ spin-spin coupling, that occurs
with the increasing of the magnetic field and is caused by
long-range Coulomb interaction, in particular by the second term
in Eq. (\ref{eq18}) . For $\mathbf{B}>\mathbf{B^*}$, a compression
of the orbitals appears, which reduces the overlap of the
wavefunctions exponentially. In addition, in Fig.\ref{fig1} it is
shown the behavior of $J(B)$ for some values of $E$. The fourth
term in Eq. (\ref{eq18}) reveals the dependence of $J$ with $E$.
The increasing of the electric field produces a raise on the
exchange energy. The increment of $E$ creates a displacement of
the value $\mathbf{B^*}$, in which the sign switch of $J$ takes
place. Thus, the most efficiency to tune the exchange energy  $J$
is acquired  for $E=0$.
\\
\begin{figure}[h]
\begin{center}
\includegraphics[angle=0,width=8.5 cm]{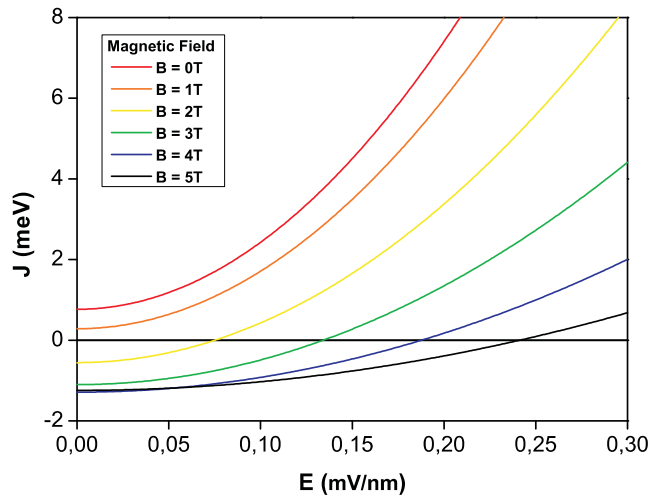}
\end{center}
\caption{Transition of exchange energy for $J(E)$. Instead of
$J(B)$ and $J(d)$, the transition in this case is from
ferromagnetic to antiferromagnetic.} \label{fig2}
\end{figure}
Fig.\ref{fig2}. shows the behavior of $J(E)$, for $B$ and $d$
fixed. Here, it is observed that it is possible to produce a sign
switch, but the transition is now from ferromagnetic $(J<0)$ to
antiferromagnetic $(J>0)$ and only with an acting $B \geq 1.35$T
field over the system. This feature yields to think in a two
quantum dots scheme, operating as a quantum gate, it could return
to its initial state without eliminating the magnetic field
interaction after the $J$ switching. If we consider that the
transition is initially from antiferromagnetic to ferromagnetic,
varying $B$ and fixing $E$ (Fig.\ref{fig3}A-B), to return to its
initial state we keep $B$ constant and increase $E$ until the
system presents a transition from ferromagnetic to
antiferromagnetic (Fig.\ref{fig3}C-D).
\\
\begin{figure}[h]
\begin{center}
\includegraphics[angle=0,width=8.0 cm]{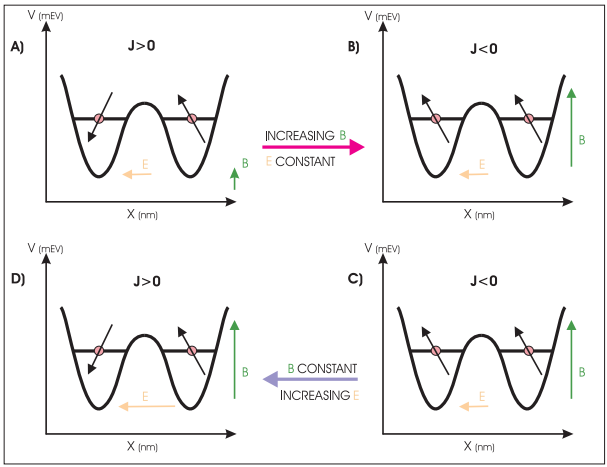}
\end{center}
\caption{Switching system for two laterally coupled quantum dots
operating as a quantum gate.}
\label{fig3}
\end{figure}
\\
The behavior of the exchange energy as a function of $d$ is showed
in Fig. \ref{fig4}. In this representation, for diverse values of
$B$, and with $E=0$, if we vary $d$ between 0 and 1.5 a transition
antiferro - ferro of $J$ is evident for $B>1T$. Similarly to Fig.
\ref{fig1}, for $d>1$, the overlapping between the wave functions
decreases exponentially. Another important characteristic which
uncover the efficiency of this model is exhibited when an increase
of the $B$ field is made, producing a diminution of the separation
distance between the dots at which $J$ changes its sign.
\\
\begin{figure}[h]
\begin{center}
\includegraphics[angle=0,width=8.5 cm]{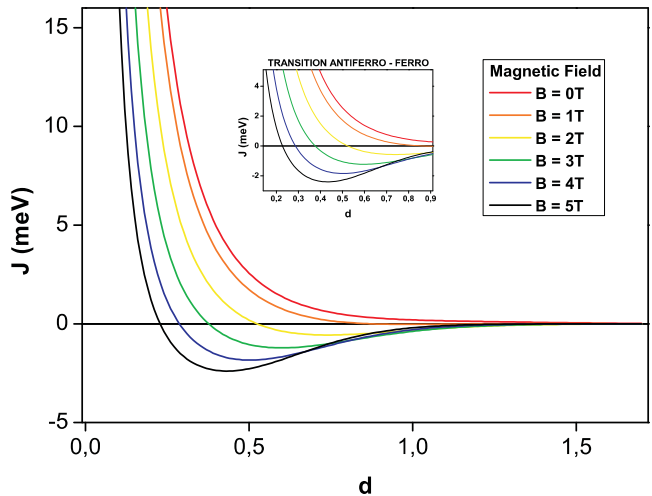}
\end{center}
\caption{Behavior of the exchange energy $J(d)$ as a function of
the quantum dots separation, keeping the electric field constant,
$E=0V$. We observe that the increasing of the magnetic field
produces a diminution of the separation distance between the dots
at which $J$ changes its sign.} \label{fig4}
\end{figure}
\\
In the last years, experiments of inelastic cotunneling for two
electrons in a single gated quantum dots have led to carry out
measurements of the exchange energy $J$\cite{zumbuhl}, showing a
high concordance with the already described in the theoretical
model.

\section{Conclusions}
We achieve a detailed description of a satisfactory model which
let us calculate the exchange energy factor $J$ analytically, for
a system of two laterally coupled quantum dots, applying the
Heitler-London formalism.
\\
The calculation of $J$ as a function of parameters such as $E$,
$B$ and $d$, and an adequate variation of them allows us to
describe a control scheme in the sign of the exchange energy,
which will help to produce qubits entanglement in the arquitecture
of Loss and DiVincenzo.
\\
Using a constant magnetic field over the two quantum dots it is
possible to switch the exchange energy sign, by means of a
changeable electric field, whose increase allows an
antiferromagnetic to ferromagnetic transition.
\\
A switching scheme of $J$ is also presented, which allows to a
quantum gate reaching its initial state after computing certain
operation, without eliminating the interactions of the electric
and magnetic fields on the system.
\vskip 1cm%
\noindent{\bf Acknowledgments}\\%

H.E.Caicedo-Ortiz thank to A.I. Figueroa for technical support in
the writing of this manuscript. S.T. Perez-Merchancano acknowledge
financial support from the \emph{Vicerrectoria de Investigaciones
- Universidad del Cauca} for his displacement to the "XII Latin
American Congress of Surface Science and its applications CLACSA –
2005".

\end{document}